\documentclass[usenatbib]{mn2e}
\usepackage[utf8]{inputenc}
\usepackage{natbib,epsfig,cite,amsmath,amssymb,graphicx,mystyle}
\usepackage{array, xcolor, caption, multirow}
\usepackage[colorlinks=true,linkcolor=blue,citecolor=blue]{hyperref}

\title[Merger rates of binary black holes and the gravitational wave background]{Metallicity-constrained merger rates of binary black holes and the stochastic gravitational wave background}
\author[I. Dvorkin, E. Vangioni, J. Silk, J.-P. Uzan, K. Olive]{Irina
Dvorkin$^{1}$\thanks{E-mail: dvorkin@iap.fr}, Elisabeth Vangioni$^{1}$, Joseph Silk$^{1,2,3,4}$, Jean-Philippe Uzan$^{1}$,\newauthor Keith A. Olive$^{5}$\\
$^{1}$Sorbonne Universit\'{e}s, UPMC Univ Paris 6 et CNRS, UMR 7095, Institut
d’Astrophysique de Paris, 98 bis bd Arago, F-75014 Paris, France \\
$^{2}$AIM-Paris-Saclay, CEA/DSM/IRFU, CNRS, Univ Paris 7, F-91191, Gif-sur-Yvette, France \\
$^{3}$Department of Physics and Astronomy, The Johns Hopkins University,
Baltimore, MD 21218, USA \\
$^{4}$BIPAC, University of Oxford, 1 Keble Road, Oxford OX1 3RH, UK \\
$^{5}$William I. Fine Theoretical Physics Institute, School of Physics and
Astronomy, University of Minnesota, Minneapolis, MN 55455, USA
}
\begin{document}

\pagerange{\pageref{firstpage}--\pageref{lastpage}} \pubyear{2016}
\maketitle
\label{firstpage}

\begin{abstract}

The recent detection of the binary black hole merger GW150914 demonstrates the existence of black holes more massive than previously observed in X-ray binaries in our Galaxy. This article explores different scenarios of black hole formation in the context of self-consistent cosmic chemical evolution models that simultaneously match observations of the cosmic star formation rate, optical depth to reionization and metallicity of the interstellar medium. This framework is used to calculate the mass distribution of merging black hole binaries and its evolution with redshift. We also study the implications of the black hole mass distribution for the stochastic gravitational wave background from mergers and from core collapse events.
\end{abstract}


\section{Introduction}

The recent detection by the Advanced Laser Interferometric Gravitational-wave
Observatory (LIGO) of the gravitational wave source GW150914
\citep{2016PhRvL.116f1102A} constitutes the first observational evidence for a
merger of a binary black hole (BBH) system. The signal matches the waveform
expected from a merger of two black holes (BHs) of masses
$36^{+5}_{-4}M_{\odot}$ and $29^{+4}_{-4}M_{\odot}$ at a luminosity distance of
$410^{+160}_{-180}$ Mpc, corresponding to a redshift of $0.09^{+0.03}_{-0.04}$
\citep{2016PhRvL.116f1102A,2016ApJ...818L..22A}. 

One of the most interesting astrophysical questions raised by this discovery is how such 'heavy' BHs can form and what are the physical conditions required. The most massive X-ray binaries with reliably measured masses reach only $20M_{\odot}$ making GW150914 the most massive stellar BBH ever observed \citep{2016ApJ...818L..22A}. The masses of remnants that can form in a supernova (SN) have been studied extensively using various techniques \citep[see][for a comprehensive review on SN explosion mechanisms]{2012ARNPS..62..407J}. One of the first quantitative approaches \citep[e.g.][]{1995ApJS..101..181W, 2008ApJ...679..639Z} was to initiate the explosion artificially in a 1-D (spherically-symmetric) stellar model. These models typically find that the final mass of the collapsed remnant is sensitive to the explosion energy, the presupernova structure, the stellar mass and the metallicity. A reverse shock is generated when the supernova front shock travels through the hydrogen envelope of the star and can decelerate a significant amount of matter, further increasing the final remnant mass. 
BH masses obtained in these models are generally below $40M_{\odot}$ with the
maximum fraction of the mass of the progenitor star winding up in the BH of up
to $\sim 25\%$ for a solar-metallicity star and up to $\sim 70\%$ for
zero-metallicity star. Note that in order to calculate the mass distribution of
BHs one needs to account for the stellar initial mass function (IMF) of the
progenitor stars which peaks at low stellar masses
\citep[e.g.][]{2001MNRAS.322..231K}. As a result, piston-driven models tend to
predict negligibly small number densities of heavy BHs. 

In recent years other models of SN collapse were built using higher-dimensional
numerical simulations and/or more accurate description of the neutrino physics
\citep[e.g.][]{2007ApJ...659.1438F,2011ApJ...730...70O,2012ApJ...747...73L,
2012ApJ...756...84M,2012ApJ...757...69U,2014ApJ...785...28K,2015ApJ...801...90P,2015ApJ...799..190C,2016ApJ...818..124E}. These studies confirmed that BHs form for progenitor
masses above $\sim 25M_{\odot}$ either via direct collapse or via fallback and
elucidated the connections between the progenitor mass and metallicity and the
remnant mass. However, it is challenging to explore a wide range of progenitor
masses and metallicites using higher-dimensional simulations due to the high
computational costs.

An alternative method is to use analytical tools to estimate the explosion energy, as was done in \citet{2006NewAR..50..492F} under the assumption that the energy reservoir is limited to the convective region bounded by the edge of the proto-neutron star and the supernova shock. This recipe was subsequently used in \citet{2012ApJ...749...91F} to study the dependence of the compact remnant mass function on the delay between core bounce and explosion and to demonstrate that the difference between possible explosion mechanisms will be detectable by gravitational wave observatories.

If the progenitors of the BBH evolve in an isolated
environment without dynamical interactions \citep[see e.g.][for an alternative scenario where BBH form through dynamical interactions in dense stellar clusters]{2014MNRAS.441.3703Z}, then the masses of the remnants as well as the merger delay time will depend also on their binary interactions. Evolutionary models \citep[e.g.][]{2010ApJ...714.1217B,2015MNRAS.451.4086S} have shown that single stars can form black holes as massive as $\sim 100M_{\odot}$ and follow-up population synthesis codes of isolated binary evolution \citep[e.g.][]{2010ApJ...715L.138B} predict the existence of BBH systems that merge within the age of the Universe. In particular, it was recently shown \citep{2016arXiv160204531B,2016arXiv160203790E} that the stellar progenitors of GW150914 had to form in a low-metallicity environment. Other models were developed that predict the formation of massive BHs under specific conditions, such as chemically homogeneous evolution of the binary \citep[e.g.][]{2016A&A...588A..50M,2016MNRAS.458.2634M} or the direct collapse of a single, fast rotating star to a binary which subsequently merges \citep[e.g.][]{2016ApJ...819L..21L,2016arXiv160300511W}.

It has also been proposed that the first stellar generation (Population III;
PopIII) which was created from zero metallicity gas is responsible for a
significant fraction or even the majority of BBH mergers observable with
gravitational wave observatories \citep[e.g.][]{1984MNRAS.207..585B,2004ApJ...608L..45B,2012A&A...541A.120K}. \citet{2014MNRAS.442.2963K} found that the
\emph{typical} mass of PopIII BBH is $\sim 30M_{\odot}$ and the detection rate is expected to be as high as $\sim 140$ events per year \citep[although note that the evolution of massive stars at low metallicity could differ substantionally from that in metal-rich environments, see][]{2015A&A...581A..15S}. However
the importance of PopIII stars as merging BBH progenitors in the context of
realistic galaxy evolution models is still debated. Recently,
\citet{2016arXiv160305655H} concluded, based on their self-consistent
cosmological semi-analytic model that there is a only a $\sim 1\%$ probability
that GW150914 is of PopIII origin and that the GW background of BBH mergers
produced by PopIII stars is small compared with other contributions at $f\simeq
25$ Hz, where the LIGO network is most sensitive. On the other hand,
\citet{2016arXiv160306921I} found that the GW background produced by PopIII
remnants can dominate other populations and would enable tight constraints on
the PopIII properties.

The purpose of this paper is to explore the merger rate of BBHs in the context of self-consistent cosmic metallicity evolution models. In particular, we focus on the differences in the expected mass distribution of merging BBHs between different models of BH formation. With more detections of coalescing BBHs by the Advanced LIGO and VIRGO observatories expected in the near future, it might be possible to constrain the SN explosion mechanism from the observed BBH mass spectrum. We also explore the stochastic gravitational wave background from merger and SN collapse events and show that it might be possible to disentangle these two contributions with observations in the range $f \gtrsim 400$ Hz.

In this work we compare two models of BH formation, the 1-D model of \citet{1995ApJS..101..181W} and the analytic description of
\citet{2012ApJ...749...91F} which give the remnant mass as a function of
progenitor mass and metallicity for a wide range of masses and metallicities.
Since we also use \citet{1995ApJS..101..181W} for the stellar yields, this first choice
constitutes a fully self-consistent description, while the model by
\citet{2012ApJ...749...91F} is based on a physically motivated explosion
mechanism and its description of BH formation is more realistic. We also explore the model developed
by \citet{2014MNRAS.442.2963K} for PopIII stars. In addition, we consider
various prescriptions for the star formation rate and IMF. While this choice of
models is far from being exhaustive, our goal here is to show how different
prescriptions can be discriminated using upcoming observations of gravitational
waves combined with constraints from reionization and metallicity measurements.

The structure of this paper is as follows. In Section \ref{sec:models} we
describe the BH formation models we employ and the chemical evolution model in
which they are embedded. In particular, we discuss constraints from observations
of metal absorption lines and optical depth to reionization on the various
prescriptions for the SFR that we use. Possible contributions from PopIII stars
are also discussed. In Section \ref{sec:mergers} we present the BBH merger rate
as a function of the BH mass expected in different models. These results are
used in Section \ref{sec:gw} to calculate the stochastic gravitational wave
background from mergers as well as from SN collapse. We conclude in Section
\ref{sec:discussion}.

\section{Binary black hole formation scenarios}
\label{sec:models}

The formation of BHs occurs at the end of the nuclear burning phase in massive
stars and can proceed via two routes. For the lower mass end of BH formation, a
meta-stable proto-neutron star (NS) is produced, followed by a formation of a BH
through accretion of the part of the stellar envelope that could not be expelled
in the supernova. Direct collapse (sometimes called failed supernova)
into a BH occurs in the case of the most massive stars. Thus, the mass of the
remnant is determined mainly by the mass of the star at the moment of collapse,
as well as the explosion energy.

Massive stars generally experience strong winds which cause them to shed a significant fraction of their envelopes during their lifetime \citep[e.g.][]{2007A&A...464L..11M,2008NewAR..52..419V,2011A&A...527A..52G}. The strength of these winds depends on metallicity: stars at lower metallicities exhibit weaker winds due to reduced opacity and easier radiation transport. Other factors that influence the relationship between the zero-age main sequence (ZAMS) mass and the stellar core mass at the time of collapse are stellar rotation \citep[e.g.][]{2009A&A...497..243D} and the microphysics of stellar evolution \citep[e.g.][]{2015MNRAS.447.3115J,2015A&A...575A..60M}. 
The mass of the BH formed after the collapse of the core depends on the strength of the supernova explosion which determines how much material is ejected. In addition, after the shocked material slows down some of it may be decelerated and fall back onto the proto-NS, adding to the remnant mass. 

In this paper we consider two models of BH formation. Model \emph{WWp} follows \citet{1995ApJS..101..181W} up to progenitor masses of $m=40M_{\odot}$.  For stars with initial masses above $40M_{\odot}$ the remnant mass is extrapolated as follows:
\begin{equation}
 \frac{m_{\rm{rem}}}{m} = A\left(\frac{m}{40M_{\odot}}\right)^\beta\frac{1}{\left(\frac{Z}{0.01Z_{\odot}}\right)^{\gamma}+1}
\end{equation}
where $Z$ is the metallicity.
This functional form was chosen so as to match the results of \citet{1995ApJS..101..181W} at $m=40M_{\odot}$ for the range of metallicities they explored and the dependence of the remnant mass on metallicity given in \citet{2015PhRvD..92f3005C} \citep[see their Figure 4, which is based on][]{2010ApJ...714.1217B}. The fiducial values of our extrapolation are $A=0.3$, $\beta=0.8$ and $\gamma=0.2$. The yields in this case are scaled from the tabulated values of \citet{1995ApJS..101..181W} so as to ensure mass conservation.

The second model we consider, called \emph{Fryer} in what follows, is based on
the calculations of remnant masses taken from \citet{2012ApJ...749...91F} (in
particular their delayed model) with the yields taken from
\citet{1995ApJS..101..181W}. This model is based on the assumption that the
explosion is powered by a convection-enhanced, neutrino-driven engine and the
explosion energy, mass loss and fallback are calculated analytically. 

For each of these models we also consider the possibility of different BH
formation scenarios from metal-poor stars. Compared with the present-day stellar
population these stars are expected to be more massive, have smaller radii for
the same mass, and less mass loss by stellar winds during their lifetime. Well
before the actual detection of GW150914, \citet{2014MNRAS.442.2963K} found that
the typical mass of black holes formed from PopIII stars that would be
merging today is $\sim 30M_{\odot}$. Inspired by this prediction, we assume that
stars below some metallicity limit $Z_{\rm{limit}}=10^{-3}Z_{\odot}$ produce
black holes according to the relation found in \citet{2014MNRAS.442.2963K}. This
prescription is based on the evolutionary models of \citet{2001A&A...371..152M}
for zero-metallicity stars, which provide the stellar radius and core mass, and
the fitting formulae from \citet{2002ApJ...572..407B} for the remnant mass.
These models are named \emph{WWp+K} and \emph{Fryer+K}, in which case this
special prescription is applied to $all$ stars below the chosen metallicity
limit. In the following we explore different values of $Z_{\rm{limit}}$.

The models described above provide the remnant mass as a function of ZAMS mass and metallicity. We then assume that the remnant is a BH for remnant masses above $2.5M_{\odot}$ and a neutron star for lower masses. While mergers of double neutron stars and BH-neutron star binaries are expected to be detectable by gravitational wave observatories, we do not include the neutron star population here and leave it to future work.

These models of BH formation are embedded in the cosmic chemical evolution model based on \citet{2004ApJ...617..693D,2006ApJ...647..773D} and \citet{2009MNRAS.398.1782R}. We assume a Salpeter stellar initial mass function (IMF) with slope $x=2.35$ in the mass range $0.1-100M_{\odot}$ \citep{1955ApJ...121..161S}. We also consider the case of $x=2.7$ \citep[which can be the case for the high-mass tail of the IMF in dense and turbulent environments, see][]{2014ApJ...796...75C} which we denote \emph{steep IMF} (we note that this is an extreme scenario). For the cosmic star formation rate (SFR) we use the functional form of \citet{2003MNRAS.339..312S}:
\begin{equation}
 \psi(z)=\nu\frac{a\exp[b(z-z_m)]}{a-b+b\exp[a(z-z_m)]}
\end{equation}
where $z$ is the redshift. Our fiducial model is a fit to the observations of luminous galaxies compiled by \citet{2013ApJ...770...57B} and complemented by high-redshift observations from \citet{2015ApJ...803...34B} and \citet{2015ApJ...808..104O}. We use the fit parameters given in \citet{2015MNRAS.447.2575V}, namely $\nu=0.178$ $M_{\odot}$yr$^{-1}$Mpc$^{-3}$, $z_m=2$, $a=2.37$ and $b=1.8$. 

An alternative way to calibrate the SFR at high redshifts is by considering the rate of gamma-ray bursts \citep[GRBs;][]{2012ApJ...744...95R,2013A&A...556A..90W,2013arXiv1305.1630K}. We use the results of \citet{2013ApJ...773L..22T} and \citet{2015ApJ...799...32B} to obtain a fit combining low-redshift galaxy luminosity data and high-redshift GRB data. This choice corresponds to Model 2 in \citet{2015MNRAS.447.2575V} with the parameters $\nu=0.146$ $M_{\odot}$yr$^{-1}$Mpc$^{-3}$, $z_m=1.72$, $a=2.8$ and $b=2.46$. This set of models is dubbed \emph{GRB-based}.

\begin{figure}
\centering
\epsfig{file=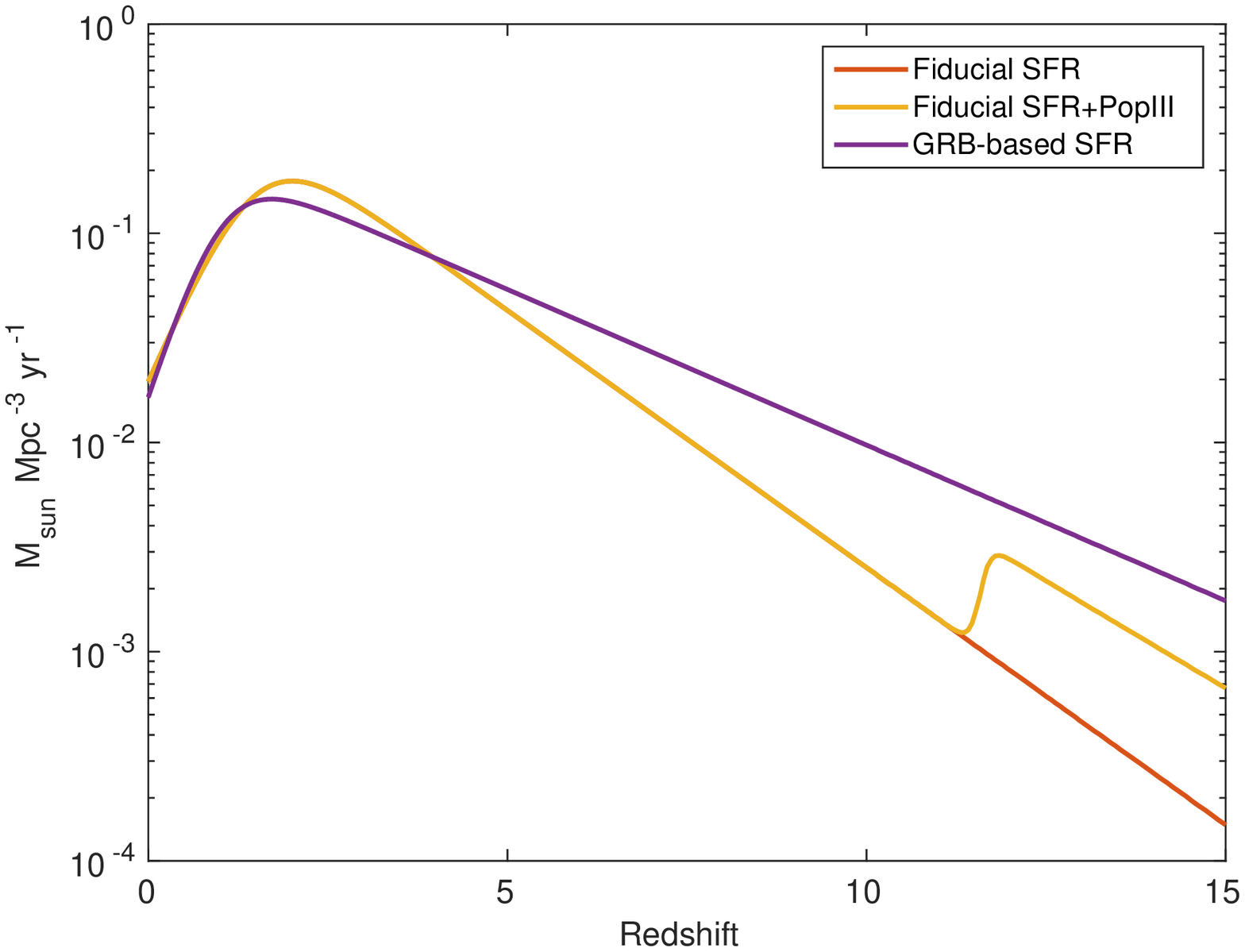, height=6.5cm}
\caption{SFR as a function of redshift for the models explored here. The
fiducial model (red line) is based on the data compilation by
\citet{2013ApJ...770...57B} and high-redshift observations from
\citet{2015ApJ...803...34B} and \citet{2015ApJ...808..104O}. Adding a PopIII
component produces a peak at $z\sim z_m$ (the combined \emph{Fiducial+PopIII}
is shown in yellow). A model that is based on high-redshift GRB data and the
normalization from \citet{2013ApJ...773L..22T} results in a higher SFR at high
redshifts (purple line).}
\label{fig:SFR}
\end{figure}

Finally, we also explore the possibility of an early PopIII component with the Salpeter IMF in the mass range $36-100M_{\odot}$ and SFR parameters $\nu=0.002$ $M_{\odot}$yr$^{-1}$Mpc$^{-3}$, $z_m=11.87$, $a=13.8$ and $b=13.36$. This set of models is named \emph{PopIII}. The SFR as a function of redshfit for a set of representative models is shown in Figure \ref{fig:SFR}. Note that the SFR calibrated to GRB data results in much higher SFR at high redshifts. 

\begin{figure}
\centering
\epsfig{file=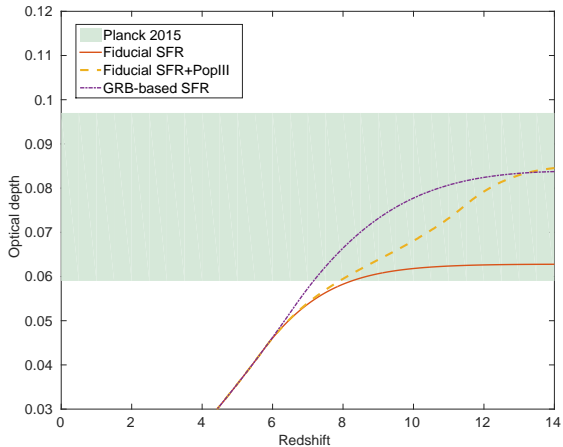, height=6.5cm}
\caption{Optical depth as a function of redshift for the set of SFR models shown
in Figure \ref{fig:SFR}, compared with the constraints from
\citet{2015arXiv150201589P} using their \emph{Planck} TT + low P combination. All the models considered here are consistent with the CMB measurement.}
\label{fig:tau}
\end{figure}

The viability of different SFR models can be studied using the constraints from
the optical depth to reionization, constrained from analysis of the cosmic
microwave background (CMB). We calculate the evolution of the volume-filling
fraction of ionized regions and the Thomson optical depth as in
\citet{2006MNRAS.373..128G}, assuming an escape fraction of $f_{\rm{esc}}=0.2$.
The number of ionizing photons for massive stars is calculated using the tables
in \citet{2002A&A...382...28S}. In Figure \ref{fig:tau} we compare the optical
depth in a set of our representative models with the value obtained by
\citet{2015arXiv150201589P}, using their \emph{Planck} TT + low P combination:
$\tau=0.078\pm 0.019$. It can be seen that all the models considered here are 
consistent with the CMB measurement. We note that the latest \emph{Planck} analysis \citep{2016arXiv160502985P,2016arXiv160503507P}, which appeared after the present article was submitted, results in a significantly lower value of the optical depth ($\tau=0.058\pm 0.012$). This new value can be accomodated if the escape fraction is lowered to $f_{esc}=0.1$, consistent with some recent measurements e.g. \citet{2016arXiv160508782M}. We intend to explore the implications of these results in future work \citep[see also the discussion in][on the effect of the escape fraction]{2015MNRAS.447.2575V}.

\begin{figure}
\centering
\epsfig{file=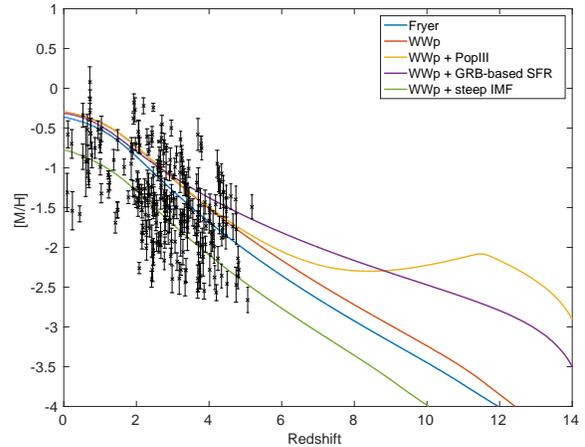, height=6.5cm}
\caption{Metallicity evolution as a function of redshift for a few representative models, compared with DLA data from \citet[][black crosses]{2012ApJ...755...89R}. Blue and red lines show the \emph{Fryer} and \emph{WWp} models, respectively, using our fiducial SFR. We also show the results for the \emph{WWp} model with alternative SFR and IMF prescriptions. At lower redshifts the different models produce very similar results, however at higher redshift the \emph{PopIII} and \emph{GRB-based} models predict much higher metallicities, which affects the masses of BHs formed in these scenarios. In the \emph{Steep IMF} model fewer massive stars are formed which leads to significantly lower metallicity (green line).}
\label{fig:metals}
\end{figure}

Various studies \citep[e.g.][]{2010ApJ...715L.138B,2012ApJ...749...91F} have
shown that the remnant mass of a given progenitor star is very sensitive to the
metallicity, mainly because stars at lower metallicity produce weaker winds. The
cosmic evolution of metallicity is therefore an important part of any model that
attempts to calculate BBH merger rates. Our model follows the chemical
enrichment of the interstellar medium (ISM) in a self-consistent manner and
reproduces the metallicities measured in high-redshift damped Ly-$\alpha$
absorbers (DLAs), as shown in Figure \ref{fig:metals}. At lower redshifts the
different models are practically indistinguishable (except for the \emph{steep IMF} model which produces much fewer massive stars), however at higher redshift the \emph{PopIII} and
\emph{GRB-based} models predict much higher ISM metallicities, a direct result
of the increased SFR in these models. Below we will explore the effect this
evolution has on the efficiency of \emph{Kinugawa}-like models. Note that the
metallicity is slightly reduced in the fiducial \emph{Fryer} model relative to
the \emph{WWp} model because in this case more mass remains locked in heavy BHs.
Indeed, as the efficiency of producing heavy BHs increases the metal yield of
the star decreases as is required by simple mass conservation.

\begin{table*}
\begin{centering}
\begin{tabular}{ |l l l l l| }
\hline

 & Model name & Ref. & Parameters & Parameter values \\ \hline
\multirow{4}{*}{BH masses} & \emph{WWp} & \citet{1995ApJS..101..181W} & $A,\beta,\gamma$ & $0.3,0.8,0.2$ \\
 & \emph{Fryer} & \citet{2012ApJ...749...91F} & - & - \\
 & \emph{WWp+K} & \multirow{2}{*}{\citet{2014MNRAS.442.2963K}} & \multirow{2}{*}{$Z_{\rm{limit}}/Z_{\odot}$} & \multirow{2}{*}{$0.001$ or $0.01$} \\
 & \emph{Fryer+K} &  &  \\ \hline
\multirow{3}{*}{SFR}  & \emph{Fiducial} & \multirow{3}{*}{\citet{2015MNRAS.447.2575V}} & \multirow{3}{*}{$\nu, z_m, a, b$} & $0.178,2.00,2.37,1.8$ \\
 & \emph{PopIII} &  &  & $0.002,11.87,13.8,13.36$ \\
 & \emph{GRB-based} &  &  & $0.146,1.72,2.8,2.46$ \\ \hline
\multirow{2}{*}{IMF} & \emph{Fiducial} & \citet{1955ApJ...121..161S} & \multirow{2}{*}{$x$} & $2.35$ \\
 & \emph{Steep IMF} & \citet{2014ApJ...796...75C} &  & $2.7$ \\
\hline
\end{tabular}
\caption{A summary of the models studied in this work. BH masses are taken either from \emph{WWp} or \emph{Fryer} and can be supplemented by \emph{Kinugawa}-like prescriptions, named \emph{WWp+K} and \emph{Fryer+K}, respectively. The \emph{PopIII} model for SFR is added to the \emph{Fiducial} model to produce a bimodal SFR as shown in Figure \ref{fig:SFR}.}
\label{tab:allmodels}

\end{centering}
\end{table*}

The various models of BH formation, SFR and IMF are summarized in Table \ref{tab:allmodels}. As we have shown, our fiducial models are consistent with the observed SFR, optical depth to reionization and cosmic metallicity evolution. 
We now proceed to calculating the birth and merger rates of BBH.

\section{Binary black hole merger rates}
\label{sec:mergers}

The direct outcome of our calculation is the birthrate of BHs as a
function of mass and redshift (or, equivalently, time) $R_{\rm{birth}}(t,m_{\rm{bh}})$ (in units of
events per unit time per unit comoving volume per unit BH mass): 
\begin{equation}
 R_{\rm{birth}}(t,m_{\rm{bh}}) = \int \psi[t-\tau(m)]\phi(m)\delta(m-g_{\rm{bh}}^{-1}(m_{\rm{bh}}))\textrm{d}m
\end{equation}
where $\tau(m)$ is the lifetime of a star of mass $m$ \citep[taken from][]{2002A&A...382...28S}, $\phi(m)$ is the IMF,
$\psi(t)$ is the SFR, $\delta(m)$ is the Dirac delta function and the ZAMS stellar
mass and black hole mass are related by some function
$m_{\rm{bh}}=g_{\rm{bh}}(m)$ which is implicit in the equation above.
$g_{\rm{bh}}$ also depends on time through its metallicity dependence and is
calculated according to the prescription of each of our models.
However the relevant quantity for gravitational waves is the merger rate which depends also on the binary fraction and the delay time between the binary formation and merger. This time delay is in most cases larger than the age of the Universe, and in general depends on the orbital parameters of the binary (semi-major axis and eccentricity) which are not resolved in our model. To circumvent this difficulty we assume a delay time distribution from the models of \citet{2016arXiv160204531B} and convolve it with the birth rate $R_{birth}(t)$ given by our models as follows: 
\begin{equation}
 R_m(t,m) = N\int_{t_{min}}^{t_{max}} R_{\rm{birth}}(t-t_d,m)P(t_d)\textrm{d}t_d\:,
 \label{eq:mrate}
\end{equation}
where $t_d$ is the delay time whose distribution is $P(t_d)\propto 1/t_d$ for
$t_{\rm{min}} < t_d < t_{\rm{max}}$ with $t_{\rm{min}}=50$ Myr and $t_{\rm{max}}$ equal to the
Hubble time. We note that, in principle, the time delay depends on the properties of the binary \citep[e.g. masses and initial orbital parameters, see][]{1964PhRv..136.1224P} as well as its environment if the binary is non-isolated \citep[e.g.][]{2014MNRAS.441.3703Z,2016PhRvD..93h4029R}. A complete treatment of the time delay distribution is beyond the scope of the present paper and will be discussed in our future work. To account for binaries that did not merge at all we normalize the
\emph{total} birth rate (sum over all masses) using the \emph{observed} rate of
$10^{-7}$ Mpc$^{-3}$ yr $^{-1}$ at $z=0$, where we used the estimate from 
\citet{2016arXiv160203842A} that assumes a power-law distribution of BH masses. This normalization is expressed
by the constant $N$. 
Note that, in principle, the delay time itself is a function of the masses of the black holes in the binary, in particular black holes with roughly equal mass are expected to merge faster, as well as those with larger total mass. The treatment of this effect is beyond the scope of the present paper and we plan to address it in future work.

\begin{figure}
\centering
\epsfig{file=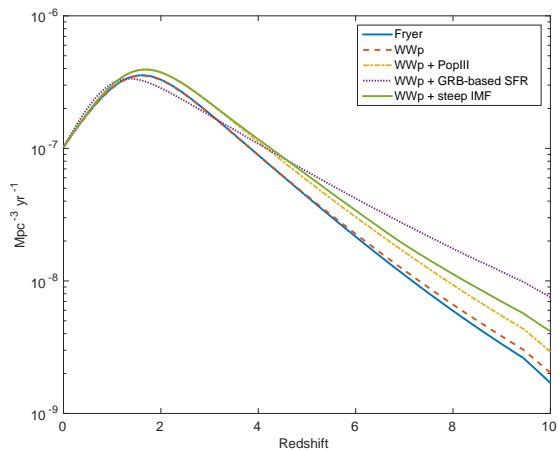, height=6.5cm}
\caption{Total merger rate of BBH as a function of redshift, normalized to
$10^{-7}$ Mpc$^{-3}$ yr $^{-1}$ at $z=0$. With the given normalization,
the curves are in practice indistinguishable below $z=3$.}
\label{fig:totmerger}
\end{figure}

The total merger rate as a function of redshift for a representative subset of
our models is shown in Figure \ref{fig:totmerger}. Note that all the models
produce very similar \emph{total} merger rates, which is a direct consequence of
the similarity in the SFR. The \emph{GRB-based} model predicts the highest BH
birth and merger rates at high redshifts due to the enhanced SFR.

\begin{figure}
\centering
\epsfig{file=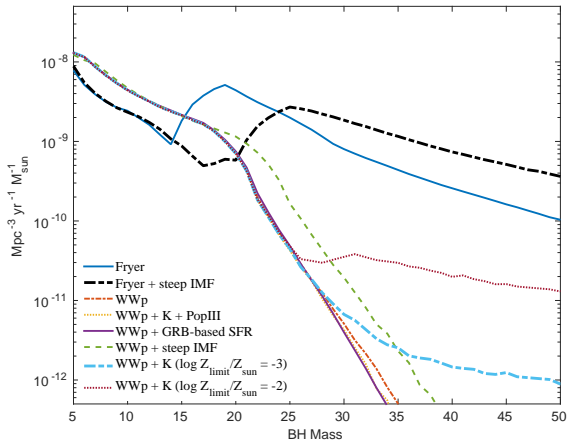, height=6.5cm}
\caption{Merger rate per unit BH mass at $z=0$ (the x-axis shows the mass of each component of the binary, assumed equal in our model). Even with the normalization of the total rate for all the models, the mass dependence is different.}
\label{fug:merger_by_mass}
\end{figure}

While it is impossible to distinguish between the different models using the
\emph{total} BBH merger rate shown in Figure \ref{fig:totmerger}, the
predictions for merger rates per unit BH mass vary among the different
prescriptions. Figure \ref{fug:merger_by_mass} shows the merger rate per unit BH
mass at $z=0$ for a few of the \emph{WWp} models, as well as two of the
\emph{Fryer} models. Here the differences between the different models become
apparent above $m_{\rm{bh}}\simeq 20 M_{\odot}$. Clearly, the \emph{Fryer} model is
much more efficient in producing massive BHs. In contrast, the fiducial
\emph{WWp} model struggles to produce massive BHs in significant amounts: these
models differ by more than two orders of magnitude for $M=30M_{\odot}$. The rate
of production of massive BHs is increased if we use steep IMF or a
\emph{Kinugawa}-like prescription for metal-poor stars. The reason for the shift towards higher BH masses in the steep IMF model is that in this case the metallicity is significantly lower, as can be seen in Figure \ref{fig:SFR}, so that even though there are fewer massive progenitor stars, the masses of the BHs that form are, on average, higher.
Note that if we include
a bimodal IMF to account for an early burst of PopIII stars (yellow dotted
line) the \emph{Kinugawa} prescription is not effective at producing BBH that
merge at $z=0$. The reason is that in this case the metallicity rises quickly at
high redshifts, as shown in Figure \ref{fig:metals} and the \emph{Kinugawa}
prescription is in fact not employed. While it produces some BBH at higher
redshift, their number density at $z=0$ is negligible, and in fact slightly
reduced relative to the fiducial \emph{WWp} model. The same effect happens when
we try to employ a \emph{GRB-based} SFR (purple dotted line overlapping with the
yellow dotted line) which increases the metallicity at high redshift.
However in the fiducial SFR model the \emph{Kinugawa} prescription significantly
increases the number of BBH mergers with masses above $\sim 30M_{\odot}$ (solid
green and dashed blue lines). As expected, the outcome of our
\emph{Kinugawa}-like models depends strongly on the assumed metallicity floor
$Z_{\rm{limit}}$, below which we employ the \emph{Kinugawa}-like prescription. When it is used for
metallicities as high as $0.01Z_{\odot}$ the difference with the \emph{Fryer}
model is less than one order of magnitude for $M\gtrsim 35 M_{\odot}$.

In contrast to the \emph{WWp} family of models, all of the \emph{Fryer} models
are very similar, except for the steep IMF model, which shifts the local peak
from $\sim 20M_{\odot}$ to $\sim 25M_{\odot}$. We note that this result might be
sensitive to the mass ranges chosen for the fits in
\citet{2012ApJ...749...91F}. 

\begin{figure}
\centering
\epsfig{file=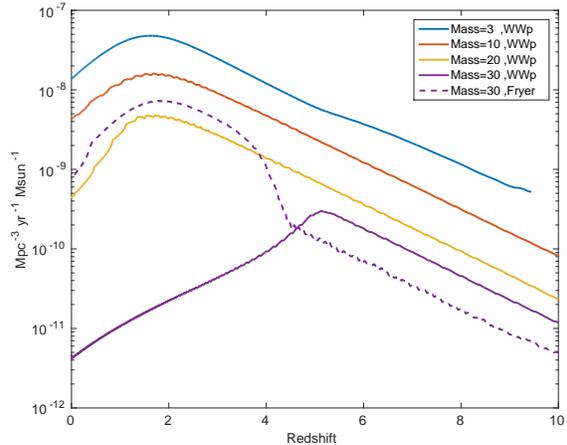, height=6.5cm}
\caption{Merger rate per unit mass as a function of redshift. Different curves correspond to $1M_{\odot}$-wide mass bins. Solid curves are for the \emph{WWp} model, while the dashed curve represents the merger rate of $30M_{\odot}$ BBH in the \emph{Fryer} model. The merger rate is dominated by low-mass binaries, with most of the contribution coming from $3M_{\odot}$ BHs (just above our chosen limit between neutrons stars and BHs). The same holds for the \emph{Fryer} model (not shown on this plot).}
\label{fig:merger_by_mass_and_z}
\end{figure}

The evolution with redshfit of the merger rate per unit BH mass is shown on
Figure \ref{fig:merger_by_mass_and_z} for the \emph{WWp} model (solid lines) and
the \emph{Fryer} model (dashed line). We note that Figure
\ref{fug:merger_by_mass} corresponds to the $z=0$ axis in Figure
\ref{fig:merger_by_mass_and_z}, and we show only $4$ mass bins to simplify the
plot. First, it can be seen that the merger rate of lower-mass BHs attains its
maximal value at around the peak of the SFR, which is not surprising given that
the chosen delay time distribution prefers very short delay times. The dominant
contribution to the overall merger rate is from small BH masses, which is the
case in all the models we considered. Events like GW150914 are therefore not
expected to constitute the majority of the merger events. Note however that in view of the sensitivity of the LIGO detector, the majority of the \emph{observed} events might well be similar to GW150914 \citep[see e.g. Figure 3 in][]{2016arXiv160204531B}. In all the models
we considered the dominant contribution was from BBHs just above our chosen limit
between neutron stars and BBHs, as a consequence of the IMF which peaks at low
masses. The \emph{WWp} model discourages the formation of BHs with masses above
$\sim 20M_{\odot}$, which only occurs at low metallicity. This explains the peak
of the mergers of $30M_{\odot}$ BHs which occurs at $z\sim 5$ in this model. For
comparison we also show the merger rate of $30M_{\odot}$ BHs in the \emph{Fryer}
model, in which case they continue to be formed and merge up to low redshfit,
with the peak occuring at $z\sim 2$. The difference between these two models in
their predictions for the merger rate of $30M_{\odot}$ BHs, shown in Figure
\ref{fug:merger_by_mass} is visible here for the whole redshift range. 

We now turn to consider the implications of our models and the differences in the obtained BBH mass distributions for the stochastic gravitational wave background.

\section{Stochastic gravitational wave background}
\label{sec:gw}

The stochastic background of gravitational waves is usually expressed in terms of the dimensionless density parameter:
\begin{equation}
 \Omega_{\rm{gw}}(f) = \frac{1}{\rho_c}\frac{\rm{d}\rho_{\rm{gw}}}{\textrm{d}\ln f}
\end{equation}
where $\rho_{\rm{gw}}$ is the gravitational energy density and $\rho_c$ is the critical energy density of the Universe. The density parameter is given by \citep{2011RAA....11..369R}:
\begin{equation}
 \Omega_{\rm{gw}}(f_o) = \frac{8\pi G}{3c^2H_0^3}f_o\int d\theta p(\theta)\int
dz
\frac{R_{\rm{source}}(z,\theta)}{(1+z)E_V(z)}\frac{\textrm{d}E_{\textrm{gw}}
(\theta)}{\textrm{d} f}
 \label{eq:omegagw}
\end{equation}
where $f_o$ is the observed frequency, $f=(1+z)f_o$ is the frequency at
emission, $p(\theta)$ is the distribution of source parameters $\theta$ (such as
the source type, binary orbital parameters etc.), $R_{\rm{source}}(z,\theta)$ is
the source rate density, $E_V(z)=\sqrt{\Omega_{\rm{m}}(1+z)^3+\Omega_{\Lambda}}$
is the cosmological volume parameter and $\textrm{d}E_{\textrm{gw}}/\textrm{d}f$
is the gravitational spectral energy emitted.

The first contribution we consider is from the coalescence of two black holes and reflects the \emph{merger rate} of BBH computed above. The merger event can be decomposed into three phases: inspiral, merger and ringdown, for which the spectrum is given approximately by \citep{2011ApJ...739...86Z}:
\begin{equation}
\frac{\textrm{d}E_{GW}}{\textrm{d}f_e}=\frac{(G\pi)^{2/3}M_c^{5/3}}{3}
 \begin{cases}
    f_e^{-1/3},&  f_e\leq f_1\\
    \omega_1 f_e^{2/3},  & f_1 < f_e < f_2\\
    \omega_2 \left(\frac{f_e}{1+\left(\frac{f_e-f_2}{\sigma/2} \right)^2} \right)^2, & f_2 \leq f_e\leq f_3
\end{cases}
\end{equation}
where $M_c=(m_1m_2)^{3/5}/(m_1+m_2)^{1/5}$ is the chirp mass. The set of
parameters $(f_1,f_2,f_3,\sigma)$, where $f_1,f_2$ correspond to the end of the
inspiral and merger phases, respectively, is taken from
\citet{2008PhRvD..77j4017A} for the case of non-spinning BHs for each set of
masses (which we assume to be always equal). The constants $\omega_1=f_1^{-1}$
and $\omega_2=f_1^{-1}f_2^{-4/3}$ are chosen to make
$\textrm{d} E_{\textrm{gw}}/\textrm{d} f$ continuous.

The second contribution is from a collapse of a single star which reflects the BH \emph{birth rate}. We assume, following \citet{2015PhRvD..92f3005C}, that most of the energy is dissipated via the ringdown of the $\ell = 2$ dominant quasi-normal mode whose frequency is given by \citep{1989PhRvD..40.3194E}:
\begin{equation}
 f_*=\frac{\Delta(a)}{m_{\textrm{bh}}}
 \label{eq:fstar}
\end{equation} 
where $m_{\rm{bh}}$ is the mass of the BH and 
\begin{equation}
 \Delta(a) = \frac{c^3}{2\pi G}\left[1-0.63(1-a)^{0.3} \right]
\end{equation}
is a function of the dimensionless spin factor $a$. The energy spectrum of a single source is then given by:
\begin{equation}
 \frac{\textrm{d} E_{GW}}{\textrm{d} f}=\epsilon m_{\rm{bh}}c^2\delta(f-f_*)
 \label{eq:dedf_collapse}
\end{equation}
where $\epsilon$ is the efficiency of GW production. We note that there could be other contributions which depend on the details of the post-core-bounce evolution of the SN, see e.g. \citet{2013ApJ...766...43M,2013ApJ...768..115O}.

\begin{figure}
\centering
\epsfig{file=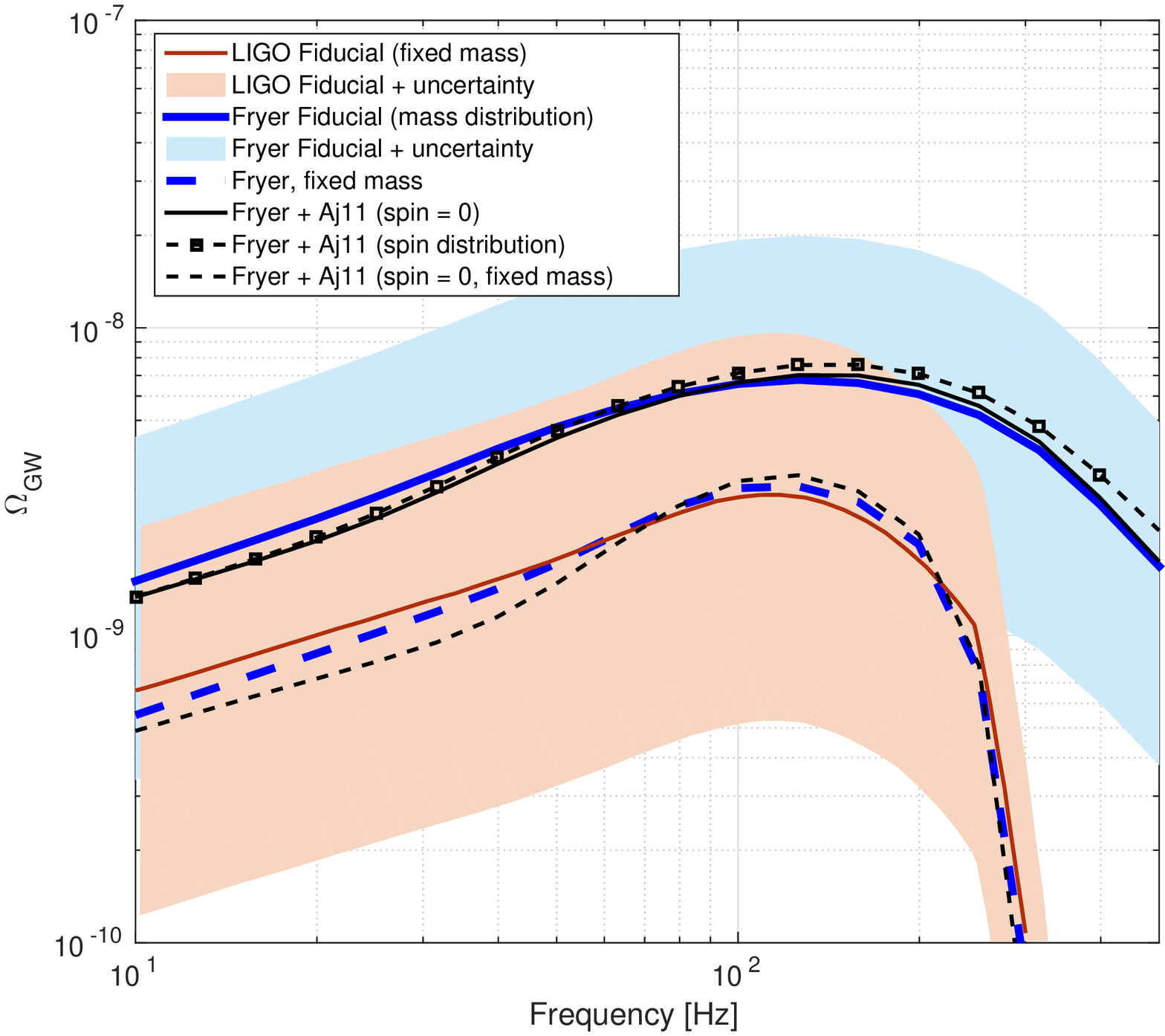, height=7cm}
\caption{Gravitational wave stochastic background from merger events. The brown curve and the light orange area are taken from \citet{2016arXiv160203847T}, where it was assumed that all the BBH have masses identical to GW150914 and the waveforms from \citet{2011PhRvD..84h4037A} were used. Dashed blue line is our calculation with the \emph{Fryer} model under the same assumption and using the waveforms from \citet{2008PhRvD..77j4017A} which produces a nearly identical result. The blue solid line corresponds to the \emph{Fryer} model with the full mass spectrum taken into account. Note that the amplitude in this case is increased and there is additional power at high frequencies which is due to low-mass BHs. The light blue shaded area corresponds to the \emph{Fryer} model taking into account the uncertainty in the measured local merging rate to which our calculation is normalized. The set of black curves shows our calculation using the waveforms from \citet{2011PhRvL.106x1101A} which include PN corrections: dashed for fixed mass, solid for a mass distribution and dashed with squares for a mass distribution and a uniform spin distribution.}
\label{fig:gw_back_ligo}
\end{figure}

Then the density parameter of gravitational waves can be calculated by plugging eqs. (\ref{eq:fstar}-\ref{eq:dedf_collapse}) into eq. (\ref{eq:omegagw}) and is given by:
\begin{equation}
 \Omega_{\rm{gw}}(f_o) = \epsilon\frac{8\pi G}{3H_0^3}\int
\frac{R_{\rm{birth}}(z',m_{\rm{bh}})}{(1+z')E_V(z')}m_{\textrm{bh}}\textrm{d} m_{\textrm{bh}}
\end{equation}
where $z'$ is a function of $m_{\rm{bh}}$. The rate of formation of BHs per unit time, per unit volume and per unit mass $R_{\rm{birth}}(z',m_{\textrm{bh}})$ is taken from our model described above where
\begin{equation}
 z' = \frac{\Delta(a)}{m_{\textrm{bh}}f_o}-1
\end{equation}
and we take the efficiency parameter $\epsilon=10^{-5}$ from \citet{2015PhRvD..92f3005C}. We assume a constant spin parameter $a=1$ for all the BHs.
The resulting stochastic gravitational wave background is depicted in Figure
\ref{fig:gw_back_ligo} (blue solids curve) calculated with the \emph{Fryer}
model, in particular using the BBH mass spectrum shown partly in
Figures \ref{fug:merger_by_mass} and \ref{fig:merger_by_mass_and_z}. The light blue shaded region corresponds to the upper and lower limits of
the local merger rate, to which we normalize the total merger rate (the factor
$N$ in eq. (\ref{eq:mrate})): $1.02^{+1.98}_{-0.79}$ $10^{-7}$Mpc$^{-3}$
yr$^{-1}$ \citep{2016arXiv160203842A}. The solid red curve is the
\emph{Fiducial} model of \citet{2016arXiv160203847T} (taken from their Figure 1)
calculated using $\textrm{d}E_{GW}/\textrm{d}f_e$ from \citet{2011PhRvD..84h4037A} and assuming
the merger rate is proportional to the cosmic SFR below metallicity
$0.5Z_{\odot}$ and the same time delay distribution as used here. Note that in this case the normalization was taken to be $1.6^{+3.8}_{-1.3}$ $10^{-8}$Mpc$^{-3}$ \citep{2016arXiv160203842A}, since they assumed that all BBHs have masses identical to
GW150914. The blue
dashed line shows our calculation where we also assumed the same masses as in
GW150914 for all the merger events, in which case our model coincides with the
LIGO \emph{Fiducial} model. The main difference between these models is the SFR:
while we use our \emph{Fiducial} SFR model, the red solid curve is calculated
using the \emph{GRB-based} model (see Figure
\ref{fig:SFR}). The difference between these models has a negligible effect on
the spectrum of the stochastic gravitational wave background. Another difference
is that in our \emph{Fryer} model with fixed mass the BBH birthrate is
in practice proportional to the SFR (since all the BHs are assumed to be born with
the same mass, the mass distributions shown in Figures \ref{fug:merger_by_mass}
and \ref{fig:merger_by_mass_and_z} are irrelevant). The LIGO \emph{Fiducial}
model, on the other hand, assumes that the birthrate is proportional to the SFR
at metallicity below $Z_{\odot}/2$, which however is proportional to the total
SFR for most of the cosmic history (see Figure \ref{fig:metals}). Since the same overall normalization to the
local observed rate is used for both the red solid and black dashed curves, this
difference is also unimportant. 

On the other hand, it can be seen that if the whole mass distribution
is taken into acount the spectrum of the stochastic background shifts to higher
values, since the BBH population is dominated by low-mass binaries. It is
therefore important to correctly account for the mass distribution of BBHs
discussed above. We note that there is also significant uncertainty due to the poorly constrained time delay distribution \citep[see the discussion in][]{2016arXiv160203847T} whose detailed treatment we leave to future work.

We also explore the effect of BH spins and post-Newtonian (PN) corrections by taking the spectrum from \citet{2011PhRvL.106x1101A} (black lines in Figure \ref{fig:gw_back_ligo}). \citet{2011PhRvL.106x1101A} matched a PN description of the inspiral phase to a set of numerical relativity simulations to obtain an analytical inspiral-merger-ringdown waveform family. While PN corrections to the waveform are crucial in parameter estimation of merger events \citep[e.g.][]{2006LRR.....9....4B}, their contribution to the stochastic background is small both in the case of fixed masses (dashed black line) and a distribution of BH masses (solid black line). Moreover, adding a uniform distribution of BH spins in the range $[-0.85,0.85]$ (the range of validity of the models of \citet{2011PhRvL.106x1101A}; black dashed line with squares) does not significantly affect the result.

\begin{figure}
\centering
\epsfig{file=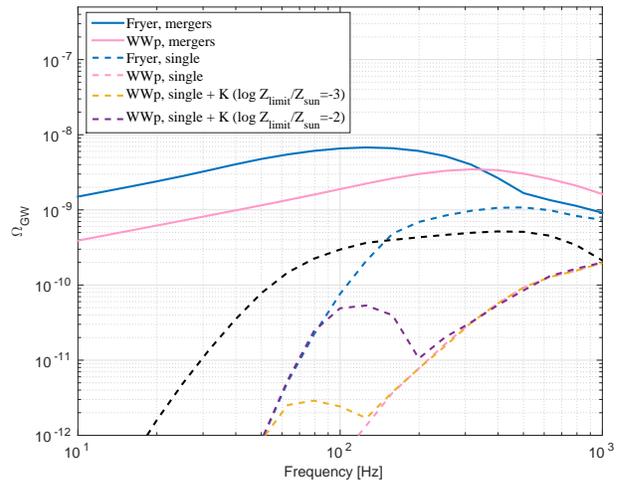, height=7cm}
\caption{Gravitational wave stochastic background from single star collapse
(dashed) and from merger events (solid). Blue curves correspond to the
\emph{Fryer} model (the blue solid line is the same as in the previous Figure),
whereas the pink lines correspond to the fiducial \emph{WWp} model. As expected,
the \emph{Fryer} models predict higher amplitude and lower frequency of the peak
owing to the larger BH masses. Note that for the \emph{Fryer} models the
contributions from the single and binary populations are separated in frequency
and can be measured with experiments that are sensitive at $f \gtrsim 400$ Hz
(although note the caveats discussed in the text). \emph{WWp} models with a
\emph{Kinugawa}-like prescription for BH formation from metal-poor stars are
shown in yellow and green for a metallicity limit of
$Z_{\rm{limit}}=0.001Z_{\odot}$ and $Z_{\rm{limit}}=0.01Z_{\odot}$,
respectively. For comparison, we also show one of the models from \citet{2015PhRvD..92f3005C} (black dashed line, see text). All of the \emph{single} models assume a dimensionless spin of $a=1$.}
\label{fig:gw_back}
\end{figure}

Figure \ref{fig:gw_back} includes the contribution from SN collapse events
(dashed lines) as well as from mergers (solid lines) and compare several of the
models discussed above. Interestingly, in our fiducial \emph{Fryer} model (blue)
the two contributions are somewhat separated in frequency. Even though the
amplitude of the contribution from SN collapse is at present highly uncertain,
this separation might be observable with experiments sensitive in the $f\gtrsim
400$ Hz frequency domain, and when the rate of observed events becomes
sufficiently large so as to reduce the uncertainty bands. Such an observation
will provide a clear handle on the two different populations - single and binary
- of BHs. We note however that there may be additional contributions in this
frequency range, such as from binary neutron star mergers not discussed here.
Furthermore, the mechanism of generation of gravitational waves during SN
collapse is highly uncertain
\citep[see e.g.][]{2013ApJ...766...43M,2013ApJ...768..115O}.

The stochastic background from mergers and SN collapse in the fiducial \emph{WWp} is shown by the pink solid and dashed lines in Figure \ref{fig:gw_back}, respectively. The peak of the merger contribution is shifted toward higher frequency with respect to the \emph{Fryer} model due to the lower BH masses in \emph{WWp}. We also plot the background from SN collapse in \emph{WWp} models that use a \emph{Kinugawa}-like prescription for metal-poor stars (yellow and green dashed lines). This prescription produces a peak at around $150$ Hz, but its amplitude is very small owing to the small number density of such SNe. For comparison we also show (black dashed curve) the calculation of the background from SN collapse from \citet{2015PhRvD..92f3005C}, where they assumed that $50\%$ of the progenitor mass goes into the BH (their Model 1). We note that all the \emph{Fryer} models (i.e. with steep IMF, PopIII stars, GRB-based SFR) produce very similar GW background and we do not expect the differences to be observable. 

\section{Discussion}
\label{sec:discussion}

This article explores different scenarios of BH formation in the context of cosmic chemical evolution models which are consistent with measured star formation rates, optical depth to reionization and metallicity evolution of the interstellar medium. We have shown that the analytic model of \citet{2012ApJ...749...91F} is much more efficient in producing heavy BBH, such as GW150914 recently observed by Advanced LIGO, than the piston-driven model of \citet{1995ApJS..101..181W}. While the sensitivity of the BBH mass distribution to the SN collapse model is not surprising \citep[e.g.][]{2012ApJ...749...91F}, our approach provides a convenient cosmological framework for the analysis of future observations.

We investigated various SFR prescriptions and found that models that produce large amounts of stars at high redshifts, such as GRB-based SFR and bimodal SFR which includes a contribution from PopIII stars, are not favourable to heavy BH production because of the accompanying rise in metallicity. On the other hand, models with very steep IMF which produce few massive stars also result in low cosmic metallicity which leads to higher BH masses. These results demonstrate the importance of self-consistent modeling. After this paper was submitted the \emph{Planck} collaboration released new constraints on the optical depth to reionization. We intend to study the consequences of these new results in future work.

The role of PopIII stars in producing heavy BBH which merge within the age of the Universe is currently debated. Our results support the conclusion of \citet{2016arXiv160305655H} in that the contribution of PopIII remnants is sub-dominant in realistic BBH formation models. In particular, we find that the inclusion of special prescriptions of BH formation from metal-poor stars, such as the model proposed by \citet{2014MNRAS.442.2963K} does not affect the mass distribution of merging BBHs in realistic scenarios, such as our set of \emph{Fryer} models. Moreover, we calculated the stochastic gravitational wave background and found that the contribution from PopIII stars is negligible in the entire frequency range we explored ($f\sim 10-1000$ Hz).

The analysis presented in this paper is far from being exhaustive and we plan to explore other BBH formation models in future work. We expect that future detections of merging BBH will enable us to discriminate between the different models of SN collapse and BBH evolution.

In this paper we normalized the total merger rate to the observed single event and used a universal distribution of delay times from stellar birth to merger. In reality the delay time depends on the orbital parameters of each binary, and the overall normalization also depends on the binary fraction. We plan to include a realistic model of binary orbital parameters in a forthcoming paper.

Finally, we plan to apply the approach presented in this paper to a full galaxy evolution model in order to calculate the anisotropy of the stochastic gravitational wave background and the cross-correlation between the optical signal from galaxies and gravitational waves from BBH mergers.

\section*{Acknowledgments}                                                      
We thank Vuk Mandic and Chris Belczynski for their constructive comments on the manuscript and the referee Tania Regimbau for very helpful suggestions. The work of ID and JS was supported by the ERC Project No. 267117 (DARK) hosted by Universit\'{e} Pierre et Marie Curie (UPMC) - Paris 6, PI J. Silk. JS acknowledges the support of the JHU by NSF grant OIA-1124403. The work of KAO was supported in part by DOE grant DE-SC0011842 at the University of Minnesota. This work has been carried out at the ILP LABEX (under reference ANR-10-LABX-63) supported by French state funds managed by the ANR
within the Investissements d'Avenir programme under reference ANR-11-IDEX-0004-02.

\bibliographystyle{mn2e}
\bibliography{gw}

\end{document}